\newcommand{\be}{\begin{equation}}
\newcommand{\ee}{\end{equation}}
\documentstyle[12pt]{article}
\newlength{\dinwidth}
\newlength{\dinmargin}
\begin{document}

\title{
\hfill{\large INP Report 1666/PH (1994)}\\
\vspace{1 cm}
       {\bf Gluons from logarithmic slopes of $F_{2}$
            in the NLL  approximation}
\vspace{1cm}
\author{
       K.Golec--Biernat \thanks{e--mail address:  golec@bron.ifj.edu.pl}
\\[15pt]
{\it Department of Theoretical Physics}
\\
{\it Institute of Nuclear Physics}
\\
{\it Radzikowskiego 152, Krak\'ow, Poland}
\\
       }
\date{}
      }

\maketitle
\vspace{3cm}
\begin{abstract}
 We make a critical, next--to--leading order, study of the accuracy
 of the "Prytz" relation, which is frequently used to extract
 the gluon distribution at small $x$ from the logarithmic slopes of
 the structure function $F_{2}$. We find that the simple relation
 is not genarally valid in the HERA regime, but show that it is a reasonable
 approximation for gluons which are sufficiently singular at small $x$.
\end{abstract}

\setcounter{page} {0}
\thispagestyle{empty}


\newpage
 Recently a new method of determination of the gluon distribution in
 the nucleon from the deep-inelastic scattering (DIS) data was proposed
  \cite{prytz}.
 The method is based on an approximate relation between the gluon
 density and the logarithmic slopes of the nucleon structure function $F_{2}$,
 derived using the leading logarithmic approximation (LLA) formulae.
 The crucial point in the derivation is the observation that for the
 existing parametrizations of the parton distributions
 the logarithmic slope of $F_{2}$ at small $x$ ($x < 10^{-2}$)
 depends mostly on the gluon density and the sea quarks
 contribution can be neglected. In this case
 the following formula for the
 logarithmic slope of $F_{2}$ is valid
 (for $n_{f}=4$)
 \be
   \label{slope1}
   \frac{dF_{2}}{d logQ^{2}} \approx \frac{5 \alpha_{s}}{9 \pi}
     \int_{x}^{1} dz\ G(\frac{x}{z},Q^{2})\ P_{qg}(z)\ ,
 \ee
  where $P_{qg}(z) = (1-z)^{2}+z^{2}$ is the Altarelli--Parisi splitting
  function in LLA.

  The integral in (\ref{slope1}) was approximated
  by the value of the gluon distribution at the point $2 x$,
  and the final relation was found
  \be
    \label{slope3}
    \frac{dF_{2}}{d log Q^{2}}(x)\approx \frac{10 \alpha_{s}}{27 \pi}\ G(2x)\ .
  \ee
  to be valid at small $x$ below $10^{-2}$.

  Relation  (\ref{slope3}) was  used by the H1 colaboration at DESY
  to estimate the gluon
  distribution in the proton from the first measurement of the structure
  function $F_2$ in the DIS region at the e-p
  colider HERA at DESY\cite{h1}.
  The slope at l.h.s of relation (\ref{slope3}) was determined at
  $Q^{2} = 20~GeV^{2}$ from the data and the gluon distribution was found
  to exhibit strong  rise when $x \rightarrow 0$.

  Relation (\ref{slope3}) helps to estimate the gluon distribution
  in LLA, while most of existing predictions of the small $x$ behaviour
  of the parton distributions are done in the next-to-leading logarithmic
  approximation (NLLA).
  It means that the corrections which are
  one order higher in  $\alpha_{s}$ are taken into account both in
  the Altarelli--Parisi equations
  and in the structure functions
  to determine  the parton distributions.
  A natural question arises whether relation (\ref{slope3}) remains
  a good approximation also for NLLA gluon ditributions.
  The author of paper \cite{prytz} claims that this relation  is
  valid with an accuracy of around $20~\%$, taking into account many
  uncertainties, also those resulting from  higher order  $\alpha_{s}$
  corrections.
  We will show that this is not generally true in NLLA.
  The validity of relation (\ref{slope3})
  crucially depends on a type of gluon distribution.

  In order to show this we computed the logarithmic slope of $F_{2}$
  on l.h.s of formula (\ref{slope3}) for three different sets of existing
  parton distribution parametrizations at small $x$ ; MRS parton distributions
  $D^{'}_{-}$ and $D^{'}_{-}$ \cite{mrs} , and GRV ones \cite{grv}.
  All of them are extrapolations from the region of high $x$, where
  the fixed targed DIS data exist.
  These distributions represent the whole spectrum of possible behaviours
  of the sea quarks and gluon distributions at small $x$.
  The Lipatov motivated $D^{'}_{-}$ parametrization
  predicts strong rise of gluon and sea, proportional to $x^{-0.5}$ when
  $x \rightarrow 0$, whereas $D^{'}_{0}$ distributions are flat. They
  tend to constant values in the limit of small $x$.
  The gluon and sea distributions from the GRV set lie in between. They
  are much steeper than $D^{'}_{0}$ distributions but not as steep as
  $D^{'}_{-}$ ones.
  We computed the slope of $F_{2}$
  numerically, using a part of a computer program,
  which solves the Altarelli--Parisi equations in NLLA, prepared to analyse
  the DIS data from HERA \cite{gol}.
  Having done that we compared the computed exact slopes of $F_{2}$ to
  r.h.s of relation (\ref{slope3}), which  can easily be found for the parton
  parametrizations under consideration.

  The results of this analysis are shown in {\bf Fig.1}. We plot  there
  the exact and approximate slopes (r.h.s of relation (\ref{slope3}))
  computed at $Q^{2} = 20~GeV^{2}$,
  as well as their ratios.
  It is easy to see that the  validity of
  relation (\ref{slope3}) in NLLA depends strongly on the parton distributions.
  While for the steep gluons from the $D{'}_{-}$ set the relation is still
  approximately true, then for the flat type $D^{'}_{0}$ gluons
  it is dramatically violated. For the GRV parametrization relation
  (\ref{slope3}) could be acceptable but for $x$ bigger then $10^{-4}$.

   In order to clarify the results from {\bf Fig.1} we computed
   contributions of different terms to the exact NLLA formula for the
   logarithmic slope of $F_{2}$.
   This formula  has a form of
   (\ref{slope1}), where the NLLA splitting function $P^{(1)}_{qg}$ \cite{fur}
   was added to LLA one
   \be
   P_{qg}(z) \rightarrow P_{qg}(z) + \frac{\alpha_{s}}{2 \pi} P^{(1)}_{qg}(z)
   \ee
   In addition,
   there are also terms with quark distributions
    with LLA and NLLA contributions in the exact formula for the slope.
   In  {\bf Fig.2} we plotted
   all four contributions to $d F_{2}/d logQ^{2}$ at $Q^{2} = 20~GeV^{2}$,
   coming
   from the gluon ({\it on the left}) and quark ({\it on the right})
   terms computed in LL  ({\it dashed lines}) and NLL ({\it dotted lines})
   approximations.
   The exact slopes ({\it continuous lines}) are also shown.
   The computation was performed for the  parton distribution sets under
   consideration. We immediately
   see that the quark contributions in both approximations
   can be neglected in the derivation of the approximate relation. The slope
   of $F_{2}$ is determined mostly by the gluon terms. For the singular
   MRS($D^{'}_{-}$) gluons the LLA contribution dominates, while for the flat
   MRS($D^{'}_{0}$) ones the NLLA contibution is as important as the LLA
   one. This is the reason why the approximate relation (\ref{slope3}) is
   not valid in NLLA for the flat type gluon distributions. In such case
   relation (\ref{slope3}) applied to extract gluon from data may overestimate
   the steepness of the gluon distribution.
   We illustrate the last statement in {\bf Fig.3},
   where the exact gluons used to compute the  slope of $F_{2}$
   and the gluons estimated from relation (\ref{slope3}) are plotted.

  In conclusion, the simple relation (\ref{slope3})
  between the logarithmic slope of $F_{2}$ and the gluon distribution
  is not generally
  valid in the NLL approximation. It remains a reasonable approximation
  only for gluons which are sufficently singular at small $x$.

  Fortunately, the gluons estimated by the H1 collaboration from the first
  data seem to be  steep enough to exclude the case with flat-like  gluons.
  Nevertheless, the problem of determination of the gluon distribution
  at small $x$ is much deeper from the theoretical point of view.
  We have discussed this problem within the standard approach based on the
  Altarelli--Parisi evolution equations and the standard factorization
  formula. A different, more appropriate at small $x$ approach was proposed
  \cite{cat,ask}. It is based on the Lipatov equation and
  a new factorization theorem.
   This approach may lead in principle to different results from
   those based on the Altarelli--Parisi equations \cite{slp}.
   A future analysis of  new data from HERA should clarify these problems.

\vskip 1cm

\noindent{\Large\bf Acknowledgment}

  I would like to thank J. Feltesse,  M.W. Krasny, J. Kwieci\'nski,
  A.D.Martin and J.Turnau  for helpful comments and discussions.
  The EC "Go  West" fellowship is gratefully acknowledged.
  This work has been partially supported by the Polish Committee
  for Scientific Research under the grant 2 0198 91 01.
\newpage

\newpage

\noindent{\large {\bf Figure Captions}}

\begin{itemize}

\item[Fig.\ 1]
      Exact slopes computed in NLLA (continuous lines)
      versus slopes resulting from the
      approximate Prytz relation assumed to be still valid in NLLA
      (dashed lines). The slopes are computed for
      MRS($D^{'}_{-}$), MRS($D^{'}_{0}$) and GRV parton ditributions.
      On the right hand side the ratio of the approximate to exact
      slopes is shown. All quantities are computed at $Q^{2} = 20~GeV^{2}$.

\item[Fig.\ 2]
      Parton contributions to the exact slope $dF_{2}/d logQ^{2}$ (continuous
      lines)
      at $Q^{2} = 20~GeV^{2}$ computed  for parton distributions under
      consideration in NLLA.
      Contibution from gluons is
      shown on the left and from quarks on the right.
      Dashed and  dotted
      lines show  the leading log  and
      next--to--leading  contribution respectively .

\item[Fig.\ 3]

      Comparison of the exact gluons used to compute the exact slopes of
      $F_{2}$ in NLLA (continuous line)
      to those estimated from the approximate Prytz relation
      (dashed line) at $Q^{2} = 20~GeV^{2}$. The ratios of the approximate
       gluon distributions to the exact ones are shown on the right hand side.

\end{itemize}

\end{document}